\documentclass[prb,twocolumn,showpacs]{revtex4}
\usepackage{amsmath}
\usepackage{dcolumn}
\usepackage{graphicx}

\begin{document}

\title[Short title for running header]{Slave rotor theory of the Mott transition in the Hubbard model: a new mean field 
        theory and a new variational wave function} 
\author{Tao Li$^{1}$, Tomonori Shirakawa$^{2,4}$, Kazuhiro Seki$^{2,3}$ and  Seiji Yunoki$^{2,3,4}$}
\affiliation{$^{1}$Department of Physics, Renmin University of China,
Beijing 100872, P.R.China\\
        $^{2}$Computational Condensed Matter Physics Laboratory, RIKEN, Wako, Saitama 351-0198, Japan \\
        $^{3}$Computational Materials Science Research Team,
RIKEN Advanced Institute for Computational Science (AICS), Kobe Hyogo 650-0047, Japan \\
        $^{4}$Computational Quantum Matter Research Team,
RIKEN Center for Emergent Matter Science (CEMS), Wako, Saitama 351-0198, Japan
}
\date{\today}

\begin{abstract}
A new mean field theory is proposed for the Mott transition in the Hubbard model based on the slave rotor representation 
of the electron operator. This theory provides a better description of the role of the long range charge correlation 
in the Mott insulating state and offers a good estimation of the critical correlation strength for the Mott transition.
We have constructed a new variational wave function for the Mott insulating state based on this new slave rotor mean field
theory. We find this new variational wave function outperforms the conventional Jastrow type wave function with
long range charge correlator in the Mott insulating state. It predicts a continuous Mott transition with non-divergent quasiparticle mass at the 
transition point. We also show that the commonly used  on-site mean field decoupling
for the slave rotor corresponds to the Gutzwiller approximation for the Gutzwiller projected wave function with only 
on-site charge correlator, which can not describe the Mott transition in any finite dimensional system.
\end{abstract}

\pacs{}

\maketitle
The study of Mott transition and the possible quantum spin liquid ground state around the Mott transition 
point in Hubbard type models has attracted a lot of attention in the strongly correlated 
electron system community\cite{Imada}. The organic compound with the formula 
$\kappa-(\mathrm{ET})_{2}\mathrm{Cu}_{2}(\mathrm{CN})_{3}$ 
forming a triangular lattice is particular interesting in this respect\cite{Kanoda,Kanoda0,Kanoda1}. 
It is generally believed that the 
multi-spin exchange process in the intermediate correlation regime may be crucial for the stabilization of the 
quantum spin liquid ground state\cite{Motrunich}. Measurements on 
$\kappa-(\mathrm{ET})_{2}\mathrm{Cu}_{2}(\mathrm{CN})_{3}$  seems to support such an expectation\cite{Kanoda0,Kanoda1}. 
More recently, the possibility of quantum spin liquid ground state around the Mott transition has also been 
hotly discussed for Hubbard models with a Dirac-type electron dispersion\cite{Meng,Yunoki}.

The study of the spin liquid state in Hubbard models is more complicated than in 
Heisenberg models as a result of the added complexity of the charge degree of freedom. Depending
on the value of $U$ for charge gap opening, the Hubbard model can exhibit different physics in the 
intermediate coupling regime. The spin liquid phase is possible only when the charge gap opens before
other symmetry breaking transition. Thus, an accurate estimation of the Mott transition point is 
important in the study of spin liquid phase in Hubbard models. The Jastrow-type variational wave function 
is a commonly used way to describe such an non-symmetry breaking transition. The Jastrow wave function 
is composed of the product of the Slater determinant $|\Psi_{\mathrm{FS}}\rangle$ for the free electron and a Jastrow factor describing 
the charge correlation in the system. Previous studies find that to describe the Mott transition 
from the metallic phase, the Jastrow factor must be long ranged\cite{Tosatti}. In particular, the long
range charge correlation is responsible for the binding of the holon(empty site) and the doublon(doubly 
occupied site) in the Mott insulating state. While physically appealing, there is no microscopic justification 
for the use of the heuristic form of the Jastrow factor, except in one dimension, when the collective charge 
fluctuation is the only important correction at low energy beyond the free fermion ground state\cite{Tayo}. 
At the same time, a very large system is needed to see clearly the signature of Mott transition in the Jastrow
wave function, which is governed by the long wavelength physics.

The slave rotor representation of the electron operator is an economic way to describe the 
charge degree of freedom of the Hubbard model\cite{Georges,Georges1} and is widely used in the study of the 
Mott transition of the Hubbard systems\cite{Georges,Georges1,Lee,Zhao,Senthil}. Different estimations on the 
critical value $U_{c}$ of the Mott transition have been made using various 
kind of mean field treatment of the slave rotor. In the simplest on-site mean field approximation,
the Mott transition is approached when both the coherent weight and the band 
width of the quasiparticle excitation vanish. This is believed to be true only in the limit of infinite dimensions
, when the spatial correlation is irrelevant.  This problem can be solved by a large-N treatment of the U(1) charge rotor, 
in which the unitary constraint on the charge rotor is relaxed to a requirement on the average. However, it is not clear
to what extent the approximations adopted on the rotor are relevant for electron in the original Hubbard model,
since the mean field theory breaks the U(1) gauge symmetry inherent in the slave rotor representation. 
Especially, it is not clear what is the relation between the various mean field theories with the variational
approaches.

The purpose of this paper is to bridge the variational approach and the slave rotor theory.
In this paper, we propose a new mean field theory for the Hubbard model in the slave rotor representation. The
new mean field treats better the long range charge correlation and predicts a 
continuous Mott transition for Hubbard model with non-divergent quasiparticle mass at the transition point.
We then construct explicitly the variational wave function related to this slave rotor mean field theory 
by enforcing the rotor constraint on the slave rotor mean field ground state. 
The so constructed wave function can be written as the product of a permanent for the charge sector
and a Slater determinant for the spin sector. We have performed variational Monte Carlo simulation on this new 
wave function and find that it outperforms the Jastrow wave function with long range correlator in the Mott insulating state, 
although it involves two variational parameters. At the same time, we find that the widely used on-site mean field 
approximation for the slave rotor generates the simple Gutzwiller wave function of the form $g^{D}|\Psi_{\mathrm{FS}}\rangle$
, which is not suitable for the description of Mott transition in finite dimensional systems.

In the slave rotor representation, the electron operator is written as
\begin{equation}
c_{i,\sigma}=e^{-i\theta}f_{i,\sigma}.
\end{equation}
Here $e^{-i\theta}$ is the lowering operator of a U(1) rotor that describes the charge degree of freedom of the electron,
and $f_{i,\sigma}$ is the fermionic  spinon operator that describes the spin degree of freedom of the electron. 
To recover the correct Hilbert space and the algebras among $c_{i,\sigma}$ and $c_{i,\sigma}^{\dagger}$, the rotor and the spinon degree of 
freedom should be subjected to the following constraint
\begin{equation}
L_{i}=\sum_{\sigma}f^{\dagger}_{i,\sigma}f_{i,\sigma}-1.
\end{equation}
Here $L_{i}$ is the angular momentum of the slave rotor on site $i$ and is conjugate to phase variable $\theta_{i}$.

We consider the Mott transition of the Hubbard model of the form
\begin{equation}
H=-t\sum_{<i,j>,\sigma}c^{\dagger}_{i,\sigma}c_{j,\sigma}+U\sum_{i}n_{i,\uparrow}n_{i,\downarrow},
\end{equation}
here $n_{i,\uparrow}=c^{\dagger}_{i,\uparrow}c_{i,\uparrow}$
, $n_{i,\downarrow}=c^{\dagger}_{i,\downarrow}c_{i,\downarrow}$. In the slave rotor representation,
 up to a constant and a shift in chemical potential, the model can be written as
\begin{equation}
H=-t\sum_{<i,j>,\sigma}f^{\dagger}_{i,\sigma}f_{j,\sigma}e^{i(\theta_{i}-\theta_{j})}+\frac{U}{2}\sum_{i}L^{2}_{i}.
\end{equation}
Here we have exploited the rotor constraint to rewrite the interaction term as the kinetic energy of the slave rotors.
The form of the Hamiltonian is inviting to decouple the spinon degree of freedom and the rotor degree of freedom.
This results in the following effective Hamiltonian for the rotor and the spinon
\begin{equation}
H_{s}=-t_{f}\sum_{<i,j>,\sigma}f^{\dagger}_{i,\sigma}f_{j,\sigma}
\end{equation}
and 
\begin{equation}
H_{r}=-J\sum_{<i,j>,\sigma}e^{i(\theta_{i}-\theta_{j})}+\frac{U}{2}\sum_{i}L^{2}_{i},
\end{equation}
respectively, where $t_{f}=t<e^{i(\theta_{i}-\theta_{j})}>$ and $J=t<\sum_{\sigma}f^{\dagger}_{i,\sigma}f_{j,\sigma}>$. 
After such a decoupling, the spinon part becomes a free fermion systems and has the same Hamiltonian as the free electron.
It can then be shown that at the zero temperature $J=\frac{K}{ZN}$, in which $K$ is the absolute value of the 
ground state energy of the non-interacting system on the same lattice, $Z$ and $N$ are the coordinate number and the 
total number of sites of the lattice respectively.

The rotor part is still nontrivial and further 
approximation is needed to solve it. In the commonly used on-site mean field approximation, 
one takes $x=<e^{i\theta_{i}}>$ as a site-independent constant. The decoupled rotor Hamiltonian then becomes the 
sum of independent rotors
\begin{equation}
H_{\theta}=-Kx\sum_{i}\cos\theta_{i}+\frac{U}{2}\sum_{i}L^{2}_{i}.
\end{equation}
This model can exhibit two phases depending on the value of $\frac{U}{K}$. In the small $U$ limit, the rotor will break 
the U(1) rotational symmetry and $x$ will be nonzero. This corresponds to a metallic phase in which both the 
spinon band width and the quasiparticle weight are renormalized by a factor $x$. On the other hand, when $U$ is large 
enough, the rotor will recover the U(1) rotational symmetry and $x$ will become zero. In such a case, both the spinon 
band width and the quasiparticle weight will be zero. The critical value $U_{c}$ for such a transition can be obtained 
from the self-consitent equation $x=<e^{i\theta>}$ and is given by $U=2K$. On the triangular lattice this corresponds to
$U_{c}\approx4.7t$.

The on-site mean field treatment can provide a rough understanding of the Mott transition on the triangular lattice. 
However, the critical value $U_{c}$ for the Mott transition predicted by it is much smaller than that is generally expected. 
The divergence of the spinon effective mass at the Mott transition is also not generally accepted for a finite dimensional
system. Another way to see the insufficiency of the on-site mean field approximation is to study the related 
variational wave function, which can be constructed by enforcing the rotor constraint on the mean field ground state. 
The mean field wave function of the system in the rotor representation is given by 
\begin{equation}
|\mathrm{MF}> =\prod_{i} \left(\sum_{m_{i}}\phi_{m_{i}} |m_{i}\rangle\right)  |f-\mathrm{FS} >,
\end{equation}
in which $|f-\mathrm{FS}\rangle$ denotes the spinon Fermi sea, $\phi_{m_{i}}$ is the rotor wave function on site $i$. 
The true wave function of the system is given by $|\Psi> =\prod_{i}\mathrm{P}_{i}|\mathrm{MF}\rangle$, 
in which $\mathrm{P}_{i}$ is the projector enforcing the rotor constraint Eq.(2) on site $i$.
Since the spinon occupation number on a given site can only be $1$,$0$ or $2$, the rotor angular momentum $m_{i}$ can 
only be $0$, $1$ or $-1$. In such a case, $|\Psi\rangle$ reduces to the 
usual Gutzwiller projected wave function of the form $g^{D}|\Psi_{\mathrm{FS}}\rangle$, with the factor $g$ given by 
$g=\phi_{\pm 1}/\phi_{0}$. Here $D$ is number of doubly occupied sites in the system.
As is well known, the simple Gutzwiller projected wave function can not describe 
the Mott transition in any finite dimension. Only in the limit of infinite dimensions, when the Gutzwiller projector 
can be treated exactly with the Gutzwiller approximation, the Gutzwiller wave function can predict a Mott transition
at finite $U$. At the same time, the divergence of spinon effective mass is realized only in the limit of infinite
dimensions, when all spatial correlation can be neglected.
We thus conclude the on-site slave rotor mean field approximation 
is equivalent to the commonly used Gutzwiller approximation and can not describe the Mott transition 
in finite dimensions. 

The reason for the failure of the simple Gutzwiller wave function to describe the Mott transition in finite dimension
is that the long range charge correlation, which has been proved to be crucial for a correct theory of Mott transition, 
is not properly accounted for in such a local treatment. The main effect of such long range charge correlation is 
to introduce attraction between the holon(empty site) and the doublon(doubly occupied site) in the singly occupied 
background. In the Mott insulating state, the holon and doublon will be bounded together because of such attraction. 
For smaller $U$, the holon-doublon pair will disassociate and the system will become metallic. To restore such     
long range charge correlation in the variational wave function, various kinds of charge correlator have been 
proposed. In particular, the long range Jastrow factor has been extensively adopted in the study of the Mott
transition in Hubbard models. However, the Jastrow factor is constructed from a two-body consideration 
and is heuristic in nature. In the following, we propose a new charge correlator based on a new type of slave rotor 
mean field theory. The new mean field theory overcomes the drawbacks of the on-site rotor mean field theory
and predicts a non-divergent effective mass at the Mott transition. The $U_{c}$ predicted by the new theory is 
also more reasonable.

To formulate the new mean field theory, we introduce a boson representation of the rotor degree of freedom. 
As a result of the rotor constraint, the angular momentum of the rotor can only be $0$ or $\pm1$. 
Here we introduce three boson operators $b_{0}$ and $b_{\pm1}$ to represent these three states. 
The three bosons are thus subjected to the constraint of
\begin{equation}
\sum_{\alpha}b^{\dagger}_{\alpha}b_{\alpha}=1,
\end{equation}
in which $\alpha=0,\pm 1$. We note that the three bosons carry different charges. More specifically, $b_{i,\alpha}$ carries
charge $\alpha$ relative to the singly occupied background. 
With these boson operators, the rotor Hamiltonian can be written as
\begin{equation}
H_{r}=-J\sum_{<i,j>}(s_{i}^{+}s_{j}^{-}+s_{i}^{-}s_{j}^{+})
+\frac{U}{2}\sum_{i} (n_{i,1}+n_{i,-1}),
\end{equation}
in which $n_{i,\alpha}=b^{\dagger}_{i\alpha}b_{i,\alpha}$, \  $s_{i}^{+}=b_{i,1}^{+}b_{i,0}+b_{i,0}^{+}b_{i-1}$,
\  $s_{i}^{-}=b_{i,-1}^{+}b_{i,0}+b_{i,0}^{+}b_{i,1}$.

In the large $U$ limit, most sites will be singly occupied and we expect both $n_{-1}$ and $n_{1}$ to be small.
We thus assume $b_{0}$ condenses and treat it as a c-number to be determined self-consistently from the constraint. 
The rotor Hamiltonian then becomes
\begin{eqnarray}
H_{r}=&-&J\eta \sum_{<i,j>}(b_{i,1}^{+}b_{j,1}+b_{i,-1}^{+}b_{j,-1}+h.c.)\nonumber\\
   &-&J\eta\sum_{<i,j>}(b_{i,1}^{+}b_{j,-1}^{+}+b_{i,-1}^{+}b_{j,1}^{+}+h.c.)\nonumber\\
&+&\frac{U}{2}\sum_{i}( n_{i,1}+n_{i,-1}).
\end{eqnarray}
in which $\eta=|\langle b_{0}\rangle|^{2}$. Here we note that the condensation of $b_{0}$ does not break the U(1) rotational 
symmetry of the rotor since it carries zero charge.

The above Hamiltonian can be diagonalized to give
\begin{equation}
H_{r}=\sum_{\mathrm{k}}\epsilon_{\mathrm{k}}(\beta_{\mathrm{k},1}^{+}\beta_{\mathrm{k},1}+\beta_{\mathrm{k},-1}^{+}\beta_{\mathrm{k},-1})+\mathrm{const}.
\end{equation}
Here $\epsilon_{\mathrm{k}}=\sqrt{\xi_{\mathrm{k}}^{2}-\Delta_{\mathrm{k}}^{2}}$, in which $\xi_{\mathrm{k}}=\frac{U}{2}-ZJ\eta\gamma_{\mathrm{k}}$ and
$\Delta_{\mathrm{k}}=ZJ\eta\gamma_{\mathrm{k}}$. 
$\gamma_{\mathrm{k}}=\frac{1}{Z}\sum_{\delta}e^{i\mathrm{k}\cdot \mathrm{\delta}}$, where
$\mathrm{\delta}$ denotes the vectors connecting nearest neighboring sites on the lattice.
The value of $\eta$ can be determined self-consistently from the constraint Eq.(9) and is given by
\begin{eqnarray}
\eta=<b^{\dagger}_{0}b_{0}>=2-\frac{1}{N}\sum_{\mathrm{k}}\frac{\xi_{\mathrm{k}}}{\epsilon_{\mathrm{k}}}.
\end{eqnarray}
The minimum energy for charge excitation is given by $\epsilon_{\mathrm{k}=0}$. The Mott transition occurs when 
this gap closes. From this requirement, we find the $U_{c}$ for Mott transition is given by $U_{c}=4ZJ\eta_{c}$. Here
$\eta_{c}$ is the value of $\eta$ at which the gap closes and is given by
\begin{equation}
\eta_{c}=2-\frac{1}{N}\sum_{\mathrm{k}}\frac{1-\gamma_{\mathrm{k}}/2}{\sqrt{(1-\gamma_{\mathrm{k}}/2)^{2}
-\gamma_{\mathrm{k}}^{2}}}.
\end{equation}
For a given lattice, $\eta_{c}$ is a mathematical constant. On the triangular lattice, this constant is found to be
$\eta_{c}\approx 0.8971$. Thus the $U_{c}$ for the Mott transition on the triangular lattice is 
$U_{c}\approx 3.5884 K\approx8.43 t$. This is a better estimation than that obtained by the local mean 
field treatment we discussed above\cite{large}. 

We now construct the variational wave function related to the new mean field theory. The mean field ground state
of the system is the product of the paired ground state of
$b_{1}$ and $b_{-1}$ boson and the Fermi sea for the spinon,
\begin{equation}
|\mathrm{MF}> =e^{\sum_{i,j}W(i,j)b^{\dagger}_{i,1}b^{\dagger}_{j,-1}}|0\rangle  |f-\mathrm{FS} >,
\end{equation}
in which $|0\rangle$ is the boson vacum,
\begin{equation}
W(i,j)=\frac{1}{N}\sum_{\mathrm{k}}\frac{\Delta_{\mathrm{k}}}{\xi_{\mathrm{k}}+\epsilon_{\mathrm{k}}}
e^{i\mathrm{k}\cdot(\mathrm{R}_{i}-\mathrm{R}_{j})}
\end{equation}
is the pair wave function for the $b_{\pm1}$ boson. In principle, we should also consider 
the condensate of the $b_{0}$ boson in the above mean field wave function. However, the $b_{0}$ 
condensate only contributes a multiplicative constant to the wave function, which only depends on the total
number of singly occupied sites.
Using the constraint Eq.(9), this multiplicative constant can be written in the form of a Gutzwiller 
factor $g^{D}$ and be absorbed in Eq.(15) by multiplying $W(i,j)$ with a factor $g$. The true electron 
wave fucntion of the system is obtained by enforcing the rotor constraint Eq.(2) and is given by
\begin{equation}
|\Psi\rangle=\prod_{i}\mathrm{P}_{i}|\mathrm{MF}> =\mathrm{Perm}[g\mathrm{W}]|\Psi_{\mathrm{FS}} \rangle.
\end{equation}
Here $|\Psi_{\mathrm{FS}}\rangle$ is the Fermi sea state of the free electron, $\mathrm{Perm}[g\mathrm{W}]$ is the permanent 
of the matrix $g\mathrm{W}$. $\mathrm{W}$ is a matrix of dimension $D$ with its matrix element 
given by $W(i,j)$, which is the pair wave function between doublon at site $i$ and holon at site $j$.

We have performed variational Monte Carlo study on the above wave function on the triangular lattice. As the 
computation of permanent of a matrix is exponentially expensive in its dimension, our calculation is limited 
to the large $U$ regime, when the number of doublon is small. In our calculation, we have used a
$12 \times 12$ cluster with periodic-antiperiodic boundary condition. There are two variational parameters, namely
$g$ and $\lambda=\frac{U}{2ZJ\eta}$, to be optimized in our wave function. To determine the 
$U_{c}$ for Mott transition, we use the mean field expression for the charge gap, which is given by 
$\epsilon_{\mathrm{k}=0}=\frac{U}{2\lambda}\sqrt{(\lambda-1)^{2}-1}$.
The gap closes when $\lambda$ approaches 2 from above. When $\lambda>2$, the pair wave function $W(i,j)$ is short ranged
and the holon and the doublon are bounded together. Fig.1 plot the evolution of the charge gap with $U/t$. The gap
exhibits a linear dependence on $U/t$ close to the Mott transition point. We note that the same behavior is also predicted 
by the large-N theory. From a linear extrapolation, we find
the charge gap closes around $U/t\approx 10.75$. This result is higher than the generally accepted value 
of 7 to 8.
\begin{figure}
\includegraphics[width=8cm,angle=0]{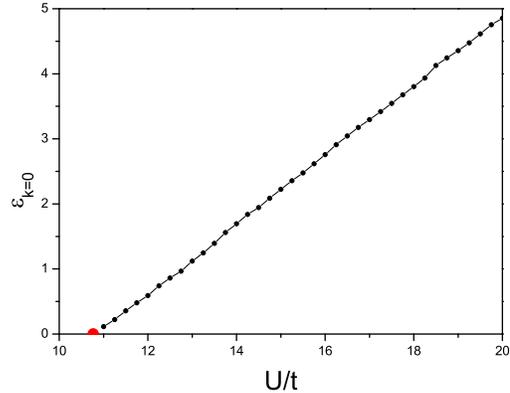}
\caption{The evolution of the charge gap with $U/t$. The red dot represent the extrapolated value of the gap closing point.
} \label{fig1}
\end{figure}

Finally, we compare our result with that produced by the wave function with long range Jastrow factor. The Jastrow wave 
function can be written as $|\Psi\rangle=e^{\sum_{i,j}v(i-j)n_{i}n_{j}}|\Psi_{\mathrm{FS}}\rangle$. Here $n_{i}$ and $n_{j}$ 
denotes the electron number at site $i$ and site $j$, $v(i-j)$ is the variational parameter introduced to control 
the charge correlation between these two sites\cite{dhdiff}. The comparison of the ground state energies 
on a $12\times 12$ cluster 
predicted by both theories is shown in Fig.2. For the Jastrow wave function, we have optimized the variational 
parameter $v(i-j)$ at all distances. The shoulder in the ground state energy around $U/t=12$ is a precursor 
of the Mott transition in the thermodynamic limit. To see clearly the signature of Mott transition in the Jastrow wave 
function, much larger lattice is needed. Compared with the Jastrow wave function, the permanent wave function 
proposed in this paper is obviously better in the insulating phase, although it involves much smaller number of
(only two) variational parameters. 
\begin{figure}
\includegraphics[width=8cm,angle=0]{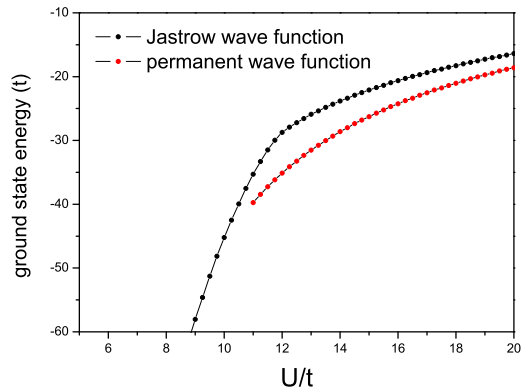}
\caption{The ground state energy as calculated from the Jastrow wave function and the permanent wave function on a 
$12\times 12$ cluster. In the former case, $v(i-j)$ at all distances are optimized.} \label{fig2}
\end{figure}

In summary, we have proposed a new mean field theory for the Mott transition in the Hubbard model based on the slave
rotor representation. The new theory can better capture the long range charge correlation, which is known to be crucial
for a correct theory of Mott transition. The new theory predicts a continuous Mott transition on the triangular lattice
around $U=8.43t$ with non-divergent quasiparticle mass at the transition point. 
The variational wave function corresponding to the 
new mean field theory has the form of the product of the permanent for the charge degree of freedom and the determinant
of the spin degree of freedom. The new wave function is found to work better than the fully optimized 
Jastrow wave function in the insulating phase. 
However, the $U_{c}$ predicted from such a wave function is still significantly larger
than the generally accepted value. We think this deficiency should be attributed to the spin degree of freedom, 
for which the superexchange interaction in the large $U$ regime is not fully accounted for by the Slater
determinant wave function. It should be noted that such superexchange effect can be incorporated in the wave function
through backflow correction in the Slater determinant\cite{Sorella,Becca}. 
We find the backflow correction can indeed improve 
significantly the estimate of $U_{c}$ in our wave function. 
A detailed discussion on this point is left as the subject of a future paper.

The computations have been done using the Magic-II Supercomputing facility in Shanghai supercomputing center, 
the RIKEN Integrated Cluster of Clusters (RICC) facility  and the RIKEN supercomputer system(HOKUSAI Great Wave). Tao Li is supported by NSFC Grant No. 11034012 and
Research Funds of Renmin University of China. This work has been also supported in part by RIKEN iTHES Project and Molecular Systems.


\begin{thebibliography}{99}
\bibitem{large}In the large-N treatment,  $U_{c}$ is given by a similar expression of the form $U_{c}=4K/\beta^{2}$ ,
with $\beta=\frac{1}{N}\sum_{\mathrm{k}}\frac{1}{\sqrt{1-\gamma_{\mathrm{k}}}}$. On the triangular lattice, it predicts that $U_{c}\approx6.16t$. However, we note that the mean field
prediction on $U_{c}$ should not be taken too seriously, since correlation effect on the spin part and the entanglement between the charge and spin part are totally ignored in such
treatment, which may shift $U_{c}$ substantially.
\bibitem{dhdiff}The correlation between two holons(or two doublons) can be different
from that between a holon and a doublon at short distance. A more accurate Jastrow wave function 
should take into account such a difference. However, we find the improvement in the ground
state energy by such a change is much smaller than the energy difference between the Jastrow and
permanent wave function.
\end{thebibliography}
\end{document}